\documentclass{llncs}
\usepackage{makeidx}  

\usepackage{amsmath}
\usepackage{color}
\usepackage{amssymb}
\usepackage{bm}
\usepackage{algorithmic,multirow}

\usepackage[ruled,vlined,linesnumbered]{algorithm2e}
\usepackage{units}
\usepackage{todonotes}
\spnewtheorem*{cor}{Corollary}{\bfseries}{\rmfamily}
\spnewtheorem*{defn}{Definition}{\bfseries}{\rmfamily}
\def\cont{\mathop{\rm cont}\nolimits}

\def\disc{\mathop{\rm disc}\nolimits}

\def\coeff{\mathop{\rm coeff}\nolimits}
\def\res{\mathop{\rm res}\nolimits}

\begin{document}
\title{The complexity of cylindrical algebraic decomposition with respect to polynomial degree}
\titlerunning{The complexity of CAD with respect to polynomial degree}  
%

\author{Matthew England\inst{1} \and James H. Davenport\inst{2}}
\authorrunning{M. England and J.H. Davenport} 
%
\tocauthor{M. England and J.H. Davenport}
\institute{
School of Computing, Electronics \& Maths, \\ Faculty of Engineering, Environment \&
Computing, \\ Coventry University, Coventry, CV1 5FB, UK \\
\email{Matthew.England@coventry.ac.uk},\\ 
WWW home page: \texttt{http://computing.coventry.ac.uk/$\sim$mengland/}
\vspace*{0.1in}
\and
Department of Computer Science, \\
University of Bath, Bath, BA2 7AY, UK\\
\email{J.H.Davenport@bath.ac.uk},\\ 
WWW home page: \texttt{http://people.bath.ac.uk/masjhd/}
}

\maketitle              

\begin{abstract}
Cylindrical algebraic decomposition (CAD) is an important tool for working with polynomial systems, particularly quantifier elimination.  However, it has complexity doubly exponential in the number of variables.  The base algorithm can be improved by adapting to take advantage of any equational constraints (ECs): equations logically implied by the input.  
Intuitively, we expect the double exponent in the complexity to decrease by one for each EC.  In ISSAC 2015 the present authors proved this for the factor in the complexity bound dependent on the number of polynomials in the input.  However, the other term, that dependent on the degree of the input polynomials, remained unchanged.

In the present paper the authors investigate how CAD in the presence of ECs could be further refined using the technology of Gr\"{o}bner Bases to move towards the intuitive bound for polynomial degree. 

\end{abstract}


\section{Introduction}
\label{SEC-Intro}

A \emph{cylindrical algebraic decomposition} (CAD) is a \emph{decomposition} of $\mathbb{R}^n$  (under a given variable ordering, so that the projections considered are $(x_1,\ldots,x_{\ell})\rightarrow(x_1,\ldots,x_k)$ for $k<{\ell}$) into cells.  The cells are arranged \emph{cylindrically}, meaning the projections of any pair with respect to the given ordering are either equal or disjoint.  In this definition \emph{algebraic} is short for semi-algebraic meaning each CAD cell can be described with a finite sequence of polynomial constraints.  A CAD is produced to be invariant for input; originally \emph{sign-invariant} for a set of input polynomials (so on each cell each polynomial is positive, zero or negative), and more recently \emph{truth-invariant} for input Boolean-valued  formulae built from the polynomials (so on each cell each formula is either true or false).

Introduced by Collins for quantifier elimination (QE) in real closed fields \cite{ACM84I}, applications of CAD included parametric optimisation \cite{FPM05}, epidemic modelling \cite{BENW06} and even motion planning \cite{SS83I}.  Some recent applications include theorem proving \cite{Paulson2012}, the derivation of optimal numerical schemes \cite{EH14} and reasoning with multi-valued functions \cite{DBEW12}.

CAD has worst case complexity doubly exponential \cite{BD07,DH88}, due to the nature of the information to be recorded rather than the algorithm used \cite{BD07}.  Let $n$ be the number of variables, $m$ the number of input polynomials, and $d$ the maximum degree (in any one variable) of the input.  Then a complexity analysis in Section 5 of \cite{EBD15} shows that the best known variant of Collins' algorithm to produce a sign-invariant CAD for the polynomials \cite{McCallum1998} has an upper bound on the size of the CAD (i.e. number of cells) with dominant term
\begin{equation}
\label{eq:BoundSI}
(2d)^{2^{n}-1}m^{2^{n}-1}2^{2^{n-1}-1},
\end{equation}
i.e. the CAD grows doubly exponentially with the number of variables $n$.

In fact, at the end of the projection stage, when we are considering $\mathbb{R}^1$, this analysis shows that we have $M$ polynomials, each of degree $D$, where $D=d^{2^{O(n)}}$ and $M=m^{2^{O(n)}}$. Of course, by replacing $\{f,g\}$ by $\{fg\}$ we can reduce $M$ at the cost of increasing $D$, but since it is much easier to find the roots of  $\{f,g\}$ than $\{fg\}$, we do not want to do so. The lower bound in \cite{DH88} shows that $D=d^{2^{\Omega(n)}}$, and that of \cite{BD07} shows that, without artificial combination, $M=m^{2^{\Omega(n)}}$. Both rely on the technique from \cite{Heintz1983}, and the formulae demonstrating this growth are not straightforward: in particular needing $O(n)$ quantifier alternations. But the underlying polynomials \emph{are} simple: all linear for \cite{BD07} and all bar two linear for \cite{DH88}. Furthermore, each polynomial only involves a bounded number of variables (generally two), independent of $n$, showing that the doubly-exponential difficulty of CAD resides in the complicated number of ways simple polynomials can interact.

An approach to improve the performance of CAD and thus this bound is to build CADs which are not sign-invariant for polynomials but truth-invariant for formulae.  This can be achieved by identifying \emph{equational constraints} (ECs): polynomial equations logically implied by formulae.  The presence of an EC restricts the dimension of the solution space and if exploited properly by the algorithm we may expect a reduction in complexity accordingly.  Intuitively, we may expect the double exponent to decrease by 1 for each independent (in a sense to be made precise later) EC available.

In \cite{EBD15} the present authors described how to adapt CAD to make use of multiple (primitive) ECs.  Suppose that our input formula consists of polynomials (as described above) and that $\ell$ suitable ECs can be identified.  The algorithm in \cite{EBD15} was shown to have corresponding upper bound dominant term
\begin{equation}
\label{eq:BoundEC1}
(2d)^{2^{n}-1} m^{2^{n-\ell} -2} 2^{\ell 2^{n-\ell} -3\ell}.
\end{equation}
So while the bound is still doubly exponential with respect to $n$, some of the double exponents have been reduced by $\ell$.  To be precise, the double exponent of $m$ (and its corresponding constant factor) is reduced while the double exponent with respect to $d$ (actually $2d$) has not.  This is due to the focus of \cite{EBD15} being on reducing the number of polynomials created during the intermediate calculations with no attempt made to control degree growth.


The present paper is concerned with how to gain the corresponding improvement to the factor dependent on $d$ to achieve the intuitive complexity bound.  The hypothesis is that this should be possible by making use of the theory of Gr\"{o}bner bases in place of iterated resultants.  
%
%
We start in Sections \ref{SUBSEC-BG}$-$\ref{SUBSEC-OtherImprovements} by reviewing background material on CAD, and then focus on CAD in the presence of ECs in Sections \ref{SUBSEC-EC}$-$\ref{SUBSEC-ECProj}.  In Section \ref{SEC-Deg} we consider how the growth of degree in iterated resultants grows compared to that of the true multivariate resultant (which encodes what is needed for CAD).  In Section \ref{SUBSEC-GB} we propose controlling this using Gr\"{o}bner Bases and in Section \ref{SEC-Example} we give a worked example of how these can precondition CAD.  In Section \ref{SEC-Complexity} we sketch how this improves upon the bound (\ref{eq:BoundEC1}) and then we finish in Section \ref{SEC-Summary} by discussing some outstanding issues.

\section{CAD with respect to equational constraints}
\label{SEC-CAD}

\subsection{CAD computation scheme and terminology}
\label{SUBSEC-BG}

We describe the computation scheme and terminology that CAD algorithms derived from Collins share.  We assume a set of input polynomials (possibly derived from input formulae) in ordered variables $\bm{x} = x_1 \prec \ldots \prec x_n$.  The \emph{main variable} of a polynomial (${\rm mvar}$) is the highest ordered variable present.  

The first phase of CAD, {\em projection}, applies projection operators recursively on the input polynomials, each time producing another set of polynomials with one less variable.  Together these define the {\em projection polynomials} used in the second phase, {\em lifting}, to build CADs incrementally by dimension.  First a CAD of the real line is built with cells (points and intervals) determined by the real roots of the univariate polynomials (those in $x_1$ only).  Next, a CAD of $\mathbb{R}^2$ is built by repeating the process over each cell in $\mathbb{R}^1$ with the bivariate polynomials in ($x_1,x_2)$ evaluated at a sample point of the cell in $\mathbb{R}^1$.  This produces {\em sections} (where a polynomial vanishes) and {\em sectors} (the regions between) which together form the {\em stack} over the cell.  Taking the union of these stacks gives the CAD of $\mathbb{R}^2$.  The process is repeated until a CAD of $\mathbb{R}^n$ is produced.  

All cells are represented by (at least) a sample point and an {\em index}.  The latter is a list of integers, with the $k$th integer fixing variable $x_k$ according to the ordered real roots of the projection polynomials in $(x_1, \dots, x_k)$.  If the integer is $2i$ the cell is over the $i$th root (counting low to high) and if $2i+1$ over the interval between the $i$th and $(i+1)$th (or the unbounded intervals at either end). 

In each lift we extrapolate the conclusions drawn from working at a sample point to the whole cell.  The validity of this approach follows from the correct choice of projection operator.  For sign-invariance to be maintained the operator must produce polynomials: {\em delineable} in a cell, meaning the portion of their zero set in the cell consists of disjoint sections; and, {\em delineable} as a set, meaning the sections of different polynomials  are identical or disjoint.  One of the projection operators used in this paper is
\begin{align}
P(B) &:= \coeff(B) \cup \disc(B) \cup \res(B).
\label{eq:P}
\end{align}
Here $\disc$ and $\coeff$ denote respectively the set of discriminants and coefficients of a set of polynomials; and $\res$ denotes either the resultant of a pair of polynomials or, when applied to a set, the set of polynomials
\[
\res(A) = \left\{\res(f_i,f_j) \, | \, f_i \in A, f_j \in A, f_j \neq f_i \right\}.
\]
We assume $B$ is an irreducible basis for a set of polynomials in which every element has mvar $x_n$.  For a general set of polynomials $A$ we would proceed by letting $B$ be an irreducible basis of the primitive part of $A$; apply the operators above; and take the union of the output with the content of $A$.  The operator $P$ was introduced in \cite{McCallum1998} along with proofs of related delineability results.

\subsection{Brief summary of improvements to CAD}
\label{SUBSEC-OtherImprovements}

As discussed in the introduction, CAD has worst case complexity doubly exponential in the number of variables.  For some problems there exist algorithms with better complexity \cite{BPR06}, however, CAD implementations remain the best general purpose approach for many.  This is due in large part to the numerous techniques developed to improve the efficiency of CAD since Collins' original work including: refinements to the projection operator \cite{Hong1990}, \cite{McCallum1998} \cite{Brown2001a}, \cite{HDX14}; the early termination of lifting, such as when sufficient for QE \cite{CH91} or for building a sub-CAD \cite{WBDE14}; and symbolic-numeric lifting schemes \cite{Strzebonski2006}, \cite{IYAY09}.  Some recent advances include further refinements to the projection operator when dealing with multiple formulae as input  \cite{BDEMW13}, \cite{BDEMW16}; local projection approaches \cite{Brown2013}, \cite{Strzebonski2014a}; decompositions via complex space \cite{CMXY09}, \cite{BCDEMW14}; and the development of heuristics for CAD problem formulation \cite{BDEW13}, \cite{EBCDMW14}, \cite{WEBD14} including machine learned approaches \cite{HEWDPB14}.

\subsection{Equational constraints}
\label{SUBSEC-EC}

As discussed in the Introduction, identifying equational constraints can improve the performance of CAD.
\begin{defn}
A {\em QFF} is a quantifier free Tarski formula: a Boolean combination ($\land,\lor,\neg$) of statements about the signs ($=0,>0,<0$) of integral polynomials.  

An {\em equational constraint} (EC) is a polynomial equation logically implied by a QFF.  An EC is said to be {\em explicit} if it is an atom of the QFF, and {\em implicit} otherwise.
\end{defn}

Collins first suggested that the projection phase of CAD could be simplified in the presence of an EC \cite{Collins1998}.  The insight is that a CAD sign-invariant for the defining polynomial of an EC, and sign-invariant for any others only on sections of that polynomial, would be sufficient.  The intuitive restriction of (\ref{eq:P}) is to use only those coefficients, discriminants and resultants which are derived from the EC polynomial, as in (\ref{eq:ECProj}) below where $F \subseteq B$ is a basis for the EC polynomial. 
\begin{align}
P_{F}(B) &:= P(F) \cup \{ {\rm res}(f,g) \mid f \in F, g \in B \setminus F \}
\label{eq:ECProj} 
\end{align}
The validity of using this operator for the first projection was verified in \cite{McCallum1999b}, with subsequent projections returning to (\ref{eq:P}).  
The operator could only be used for a single EC in the main variable of the system as the delineability result for (\ref{eq:ECProj}) could not be applied recursively, excluding its use at a subsequent projection to take advantage of any EC with corresponding main variable.  This led to the development of the operator (\ref{eq:ECProjStar}) in \cite{McCallum2001} which suffered no such reduction at the cost of including the discriminants that had been removed from (\ref{eq:P}) by (\ref{eq:ECProj}).
\begin{align}
P_{F}^{*}(B) &:= P_{F}(B) \cup \disc(B \setminus F)
\label{eq:ECProjStar}
\end{align}
See Section 2 of \cite{EBD15} for examples demonstrating these operators.
A system to derive implicit ECs was also introduced by \cite{McCallum2001}, based on the 
observation that the resultant of the polynomials defining two ECs itself defines an EC.  
This is essential for maximising the savings from ECs since the reduced operators (\ref{eq:ECProj}), (\ref{eq:ECProjStar}) are for use with a single EC; meaning the savings gained are dependent not on the number of ECs, but the number identified with different main variables.

In \cite{EBD15} the present authors reviewed the theory of reduced projection operators and deduced how it could also yield savings in the lifting phase; reducing both the number of cells we must lift over with respect to polynomials; and the number of such polynomials we lift with.  These approaches meant that the projection polynomials are no longer a fixed set (key to some CAD implementations) and that the invariance structure of the final CAD can no longer be expressed in terms of sign-invariance of polynomials.  For the worked example in \cite[Section 4]{EBD15} combining the advances in this subsection allowed a sign-invariant CAD with 1,118,205 cells to be replaced by a truth invariant CAD with 93 cells.

\subsection{CAD with ECs}
\label{SUBSEC-ECProj}

\begin{algorithm}[ht]\label{alg:ECProj}
\caption{CAD Projection using multiple stated ECs}
\SetKwInOut{Input}{Input}\SetKwInOut{Output}{Output}
\Input{A formula $\phi$ in variables $x_1,\ldots,x_n$, and a sequence of sets $\{E_k\}_{k=1}^n$; each either empty or containing a single primitive polynomial with mvar $x_k$ which defines an EC for $\phi$.
}
\Output{A sequence of sets of polynomials ready for a suitable CAD lifting algorithm.
}
\BlankLine
Extract from $\phi$ the set of defining polynomials $A_n$\label{step:Pstart}\;
\For{$k = n, \dots, 2$}{
  Set $B_k$ to the finest squarefree basis for ${\rm prim}(A_k)$\;
  Set $C$ to $\cont(A_k)$\;
  Set $F_k$ to the finest squarefree basis for $E_k$\;
  \eIf{$F_k$ is empty}{
    Set $A_{k-1} := C \cup P(B_k)$\label{step:P}\;
  }{
    Set $A_{k-1} := C \cup P_{F_i}^{*}(B_i)$\;\label{step:Pend}
  }
}
\Return $A_{1}, \dots, A_{n}; F_{1}, \dots, F_{n}$.
\end{algorithm}

Algorithm \ref{alg:ECProj} describes the CAD projection phase in the presence of multiple ECs described in the previous subsection.  Note that (as with the previous theory of multiple ECs this is based on) we assume the ECs are primitive.  Algorithm \ref{alg:ECProj} applies the best possible (smallest validated) projection operator at each stage.  The word \emph{suitable} in the output declaration means a CAD lifting phase that makes well-orientedness checks in line with the theory of McCallum's projection operators (see \cite{McCallum1998}, \cite{McCallum1999b}, \cite{McCallum2001} for details).  

Algorithm \ref{alg:ECLift} is one such suitable lifting algorithm.  It uses the $F_i$ (knowledge of which projection steps made use of an EC) to tailor its lifts: lifting only with respect to EC polynomials (steps \ref{step:L0}$-$\ref{step:L2}) and only over cells where an EC was satisfied (steps \ref{step:C0}$-$\ref{step:C2}) (lifting trivially to the cylinder otherwise).  The correctness of these algorithms was proven in \cite{EBD15}.

\begin{algorithm}[ht]\label{alg:ECLift}
\caption{CAD Lifting using multiple stated ECs}
\SetKwInOut{Input}{Input}\SetKwInOut{Output}{Output}
\Input{The output of Algorithm \ref{alg:ECProj}: two sequences of polynomials sets $A_{1}, \dots, A_{n}; F_{1}, \dots, F_{n}$, the latter subsets of the former. 
}
\Output{Either: $\mathcal{D}$, a truth-invariant CAD of $\mathbb{R}^n$ for $\phi$ (described by lists $I$ and $S$
   of cell indices and sample points); or 
{\bf FAIL}, if not well-oriented.
}
\BlankLine
%
If $F_1$ is not empty then set $p$ to be its element; otherwise set $p$ to the product of polynomials in $A_1$\label{step:Bstart}\;  
Build $\mathcal{D}_1 := (I_1,S_1)$ according to the real roots of~$p$\;
\If{$n=1$}{
\Return $\mathcal{D}_1$\;\label{step:Bend}
}
\For{$k=2, \dots, n$\label{step:Lstart}}{
  Initialise $\mathcal{D}_k = (I_k, S_k)$ with $I_k$ and $S_k$ empty sets\;
  \eIf{$F_{k}$ is empty\label{step:L0}}{
    Set $L:=B_k$\;\label{step:L1}
  }{
  Set $L:=F_k$\;\label{step:L2}
  } 
  \eIf{$F_{k-1}$ is empty\label{step:C0}}{
    Set $\mathcal{C}_a := \mathcal{D}_{k-1}$ and $\mathcal{C}_b$ empty\label{step:C1a}\;
  }{
    Set $\mathcal{C}_a$ to be cells in $\mathcal{D}_{k-1}$ with $I_{k-1}[-1]$ even\label{step:C1b}\;
    Set $\mathcal{C}_b := \mathcal{D}_{k-1} \setminus \mathcal{C}_a$\label{step:C2}\;
  }
  \For{each cell $c \in \mathcal{C}_a$}{
    \If{An element of $L$ is nullified over $c$\label{step:WO}}
    {\Return FAIL\;\label{step:fail}
    }
    Generate a stack over $c$ with respect to the polynomials in $L$, adding cell indices and sample points to $I_k$ and $S_k$\label{step:lift}\;
  }
  \For{each cell $c \in \mathcal{C}_b$}{
    Extend to a single cell in $\mathbb{R}^k$ (cylinder over $c$), adding index and sample point to $I_k$ and $S_k$\label{step:Lend}\;
  }
}
\Return $\mathcal{D}_n = (I_n, S_n)$.
\end{algorithm}

\newpage 

Table \ref{tab:P} is recreated from \cite{EBD15} and shows the growth in the number and degree of the projection polynomials when following Algorithm \ref{alg:ECProj} under the assumption that we have declared ECs for the first $\ell$ projections (so $0< \ell \leq \min(m,n)$).   Rather than the actual polynomials created the table keeps track of sets of polynomials known to have the \emph{(M,D)-property}: the ability to be partitioned into $M$ subsets, each with maximum combined degree $D$.  

\begin{table}[ht]
\caption{Projection in CAD with projection operator (\ref{eq:ECProjStar}) $\ell$ times and then (\ref{eq:P}).}  \label{tab:P}
\begin{center}
\begin{tabular}{ccc}
\textbf{Variables}    & \textbf{Number} & \textbf{Degree}                    \\
\hline
$n$          & $m$       & $d$                   \\
$n-1$        & $2m$      & $2d^2$          \\
$n-2$        & $4m$      & $8d^4$         \\
\vdots       & \vdots    & \vdots                   \\
$n-\ell$     & $2^{\ell}m$    & $2^{2^{\ell}-1}d^{2^{\ell}}$   \\
$n-(\ell+1)$ & $2^{2\ell}m^2$ & $2^{2^{\ell+1}-1}d^{2^{\ell+1}}$    \\
$n-(\ell+2)$ & $2^{4\ell}m^4$ & $2^{2^{\ell+2}-1}d^{2^{\ell+2}}$   \\
\vdots       & \vdots                      & \vdots      \\
$n-(\ell+r)$ & $2^{2^{r}\ell}m^{2^{r}}$    & $2^{2^{\ell+r}-1}d^{2^{\ell+r}}$ \\
\vdots       & \vdots                                   & \vdots            \\
1            & $2^{2^{(n-1-\ell)}\ell}m^{2^{n-1-\ell}}$ & $2^{2^{n-1}-1}d^{2^{n-1}}$  
\end{tabular}
\end{center}
\end{table}

The (M,D)-property was introduced in McCallum's thesis and was used (along with tables like Table \ref{tab:P}) to give a detailed comparison of the complexity of several different projection operators in \cite[Section 2.3]{BDEMW16}.  
The key observation is that the number of real roots in a set with the (M,D)-property is at most $MD$ (although in practice many will be in $\mathbb{C} \setminus \mathbb{R}$).  Hence the number of cells in the CAD of $\mathbb{R}^1$ is bounded by twice the product of the final entries, plus 1. 

Define $d_i$ and $m_i$ as the entries in the Number and Degree columns of Table \ref{tab:P} from the row with $i$ Variables.  Then the number of cells in the final CAD of $\mathbb{R}^n$ is bounded by 
\begin{equation}
\label{eq:BoundGeneral}
\prod_{i=1}^n \left[2m_id_i + 1\right].
\end{equation}
Omitting the $+1$s from each term will usually allow for a closed form expression of the dominant term in the bound.  

The derivation of bound (\ref{eq:BoundEC1}) from Table \ref{tab:P} was given in \cite[Section 5]{EBD15}.  It involved considering the two improvements to the lifting phase.  The first was lifting only with respect to EC polynomials; meaning that for the purposes of the bound we could set $m_i$ to 1 for $i = n, \dots, n-\ell$.  
The second was to lift trivially (to a cylinder) over those cells where an EC was false.

Denote by $(\dagger)$ the bound on the CAD of $\mathbb{R}^{n-(\ell+1)}$ given by (\ref{eq:BoundGeneral}) but with the product terminating at $n-(\ell+1)$, as there can be no reduced lifting until this point.  The lift to $\mathbb{R}^{n-\ell}$ will involve stack generation over all cells, but only with respect to the EC which has at most $d_{n-\ell}$ real roots and thus the CAD of $\mathbb{R}^{n-\ell}$ at most $[2d_{n-\ell}+1](\dagger)$ cells.
The next lift, to $\mathbb{R}^{n-\ell-1}$, will lift the sections with respect to the EC, and the  sectors only trivially.  Hence the cell count bound is
$
[2d_{n-(\ell-1)}+1]d_{n-\ell}(\dagger) + (d_{n-\ell}+1)(\dagger)
$
with dominant term $2d_{n-(\ell-1)}d_{n-\ell}(\dagger)$.  

Subsequent lifts follow the same pattern and so the dominant term (omitting the +1s) in the cell count bound  for $\mathbb{R}^n$ is
\begin{align}
\qquad &  2d_nd_{n-1} \dots d_{n-(\ell-1)}d_{n-\ell} 
\prod_{i=1}^{n-(\ell+1)}\big[2m_id_i + 1\big].
\label{eq:BoundGeneral2}
\end{align}
As shown in \cite{EBD15} using Table \ref{tab:P} (\ref{eq:BoundGeneral2}) evaluates to (\ref{eq:BoundEC1}).

\section{Controlling degree growth}
\label{SEC-Deg}

\subsection{Iterated resultant calculations}
\label{SUBSEC-IR}

As discussed in the Introduction, \cite{EBD15} showed that building truth-invariant CADs by taking advantage of ECs reduced the CAD complexity bound from (\ref{eq:BoundSI}) to (\ref{eq:BoundEC1}).  Most notably, the double exponent of the term with base $m$ (number of input polynomials) decreased by $\ell$; the number of projections made with respect to an EC.  However, the term with base $d$ (degree of input polynomials) was unchanged.  

This term is doubly exponential due to the iterated resultant calculations during projection: the resultant of two degree $d$ polynomials is the determinant of a $2d \times 2d$ matrix whose entries all have degree at most $d$, and thus a polynomial of degree at most $2d^2$.  This increase in degree compounded by $(n-1)$ projections gives the first term of the bound (\ref{eq:BoundSI}).  When building CAD in the presence of ECs many of these iterated resultants are avoided (thus reducing the \emph{number} of polynomials, but not their degree).  Indeed, the derivation of ECs via propagation is itself an iterated resultant calculation.  

The purpose of the resultant in CAD construction is to ensure that the points in lower dimensional space where polynomials vanish together are identified, and thus that the behaviour over a sample point in a lower dimensional cell is indicative of the behaviour over the cell as a whole.  Bus\'{e} and Mourrain discuss in \cite{BM09} the results of applying the iterative univariate resultant to multivariate polynomials, demonstrating decompositions into irreducible factors involving the multivariate resultants (following the formalisation of Jouanolou \cite{Jouanolou1991}).  They show that the approach will identify polynomials of higher degree than the true multivariate resultant and thus more than required for the purpose of identifying implicit equational constraints.  For example, given 3 polynomials in 3 variables of degree $d$ the true multivariate resultant has degree $\mathcal{O}(d^3)$ rather than $\mathcal{O}(d^4)$.

The key result of \cite{BM09} for our purposes follows.  Note that this, using the formalisation of resultants in \cite{Jouanolou1991} \cite[\S2]{BM09}, considers polynomials of a given \emph{total degree}.  However, the CAD complexity analysis discussed above and later is (following previous work on the topic) with regards to polynomials of \emph{degree at most} $d$ in a given variable.  For clarity we use the Fraktur font when discussing total degree and Roman fonts when the maximum degree.

\def\foo{\cite[Corollary 3.4]{BM09}}
\def\x{{\bf x}}\def\Res{\mathop{\rm Res}\nolimits}
\begin{cor}[\foo]
Given three polynomials $f_k(\x,y,z)$ of the form 
\[
f_k(\x,y,z) = \sum_{|\alpha|+i+j\le \mathfrak{d}_k}a_{\alpha,i,j}^{(k)}\x^\alpha y^iz^j \in S[\x][y,z],
\]
where $S$ is any commutative ring, then the iterated univariate resultant 
\[
\Res_y\big( \Res_z(f_1,f_2),\Res_z(f_1,f_3) \big) \in S[\x]
\]
is of total degree at most $\mathfrak{d}_1^2\mathfrak{d}_2\mathfrak{d}_3$ in $\x$, and we may express it in multivariate resultants (following the formalism of Jouanolou \cite{Jouanolou1991}) as
\begin{equation}\label{eq:BM}
\begin{array}{c}
\Res_y\big( \Res_z(f_1,f_2), \Res_z(f_1,f_3) \big) = 
(-1)^{\mathfrak{d}_1\mathfrak{d}_2\mathfrak{d}_3} \Res_{y,z}(f_1,f_2,f_3) 
\\ 
\qquad\times
\Res_{y,z,z'}\big( f_1(\x,y,z), f_2(\x,y,z), f_3(\x,y,z')\delta_{z,z'}(f_1) \big).
\end{array}
\end{equation}
Moreover, if the polynomials $f_1, f_2, f_3$ are sufficiently generic and $n > 1$, then this
iterated resultant has exactly total degree $\mathfrak{d}^2_1\mathfrak{d}_2\mathfrak{d}_3$ in $\x$ and both resultants on the right
hand side of the above equality are distinct and irreducible. \\
\rm[Although not stated as part of the result in in \cite{BM09}, under these genericity assumptions, $\Res_{y,z}(f_1,f_2,f_3)$ has total degree $\mathfrak{d}_1\mathfrak{d}_2\mathfrak{d}_3$ and 
the second resultant on the right hand side of (\ref{eq:BM}) has total degree $\mathfrak{d}_1(\mathfrak{d}_1-1)\mathfrak{d}_2\mathfrak{d}_3$.]
\end{cor}

In \cite{BM09} the authors interpret this result as follows. 
``The resultant $R_{12} := \Res_z(f_1, f_2)$ defines the projection of the intersection curve between the two surfaces $\{f_1 = 0\}$ and $\{f_2 = 0\}$. 
Similarly, $R_{13} := \Res_z(f_1, f_3)$ defines the projection of the intersection curve between the two surfaces $\{f_1 = 0\}$ and $\{f_3 = 0\}$. 
Then the roots of $\Res_y(R_{12},R_{13})$ can be decomposed into two distinct sets: the set of roots $x_0$ such that there exists $y_0$ and $z_0$ such that 
\[
f_1(x_0, y_0, z_0) = f_2(x_0, y_0, z_0) = f_3(x_0, y_0, z_0),
\]
and the set of roots $x_1$ such that there exist two distinct points $(x_1, y_1, z_1)$ and $(x_1, y'_1, z'_1)$ such that 
\[
f_1(x_1, y_1, z_1) = f_2(x_1, y_1, z_1) \qquad \mbox{and} \qquad f_1(x_1, y'_1,\allowbreak z '_1) = f_3(x_1, y'_1, z'_1).
\] 
The first set gives rise to the term $\Res_{x,y,z}(f_1, f_2, f_3)$ in the factorization of the iterated resultant
$\Res_y(\Res_{12}, \Res_{13})$, and the second set of roots corresponds to the second factor.'' Only the first set are of interest to us \emph{if} the $f_i$ are all ECs. However, for a general CAD construction, the second set of roots may also be necessary as they indicate points where the geometry of the sectors changes.

\subsection{How large are these resultants}

Suppose we are considering three ECs defined by $f_1, f_2$ and $f_3$; that we wish to eliminate two variables $z = x_n$ and $y = x_{n-1}$; and that the $f_i$ have degree at most $d$ in each variable \emph{separately}.  Then we may na\"\i{}vely set each $\mathfrak{d}_i = nd$ to bound the total degree. 

The following approach does better. Let $K=S[x_1,\ldots,x_{n-2},y,z]$ and $L=S[\xi_1,\ldots,\xi_N,y,z]$. 
Only a finite number of monomials in $x_1,\ldots,x_{n-2}$ occur as coefficients of the powers of $y$, $z$ in $f_1$, $f_2$ and $f_3$. Map each such monomial $x^\alpha=\prod_{i=1}^{n-2}x_i^{\alpha_i}$ to $\widetilde{m_j}:=\xi_j^{\max\alpha_i}$ (using a different $\xi_j$ for each monomial\footnote{It would be possible to economise: if $x_1x_2^2\mapsto\xi_1^2$, then we could map $x_1^2x_2^4$ to $\xi_1^4$ rather than a new $\xi_2^4$. Since this trick is used purely for the analysis and not in implementation, we ignore such possibilities.}) and let $\widetilde{f_i}\in L$ be the result of applying this map to the monomials in $f_i$. Note that the operation $\widetilde{\ }$ commutes with taking resultants in $y$ and $z$ (though not in the $x_i$ of course).

The total degree in the $\xi_j$ of $\widetilde{f_i}$ is the same as the maximum degree in all the $x_1,\ldots,x_{n-2}$ of $f_i$, i.e. bounded by $d$, and hence the total degree of the $\widetilde{f_i}$ in all variables is bounded by $3d$ ($d$ for the $\xi_i$, $d$ for $y$ and $d$ for $z$). If we apply (\ref{eq:BM}) to the $\widetilde{f_i}$, we see that 
$$
\Res_y\big( \Res_z(\widetilde{f_1},\widetilde{f_2}), \Res_z(\widetilde{f_1},\widetilde{f_3}) \big)
$$
has a factor $\Res_{y,z}(\widetilde{f_1},\widetilde{f_2},\widetilde{f_3})$ of total degree (in the $\xi_j$) $(3d)^3$. Hence, by inverting $\widetilde{\ }$, we may conclude $\Res_{y,z}(f_1,f_2,f_3)$ has maximum degree, in each $x_i$, of  $(3d)^3$.

The results of \cite{Jouanolou1991} \cite{BM09} apply to any number of eliminations.  In particular, if we have eliminated not $2$ but $\ell-1$ variables we will have a polynomial $\Res_{x_{n-\ell+1}\ldots x_n}(f_{n-\ell},\allowbreak \ldots,f_n)$ of maximum degree $\ell^{\ell}d^{\ell}$ in the remaining variables $x_1,\ldots,x_{n-\ell}$ as the last implicit EC.

These resultants $\Res_{x_{n-\ell+1}\ldots x_n}$ therefore only have singly-exponential growth, rather than the doubly-exponential growth of the iterated resultants: can we compute them?

\subsection{Gr\"{o}bner bases in place of iterated resultants}
\label{SUBSEC-GB}

A \emph{Gr\"{o}bner Basis} $G$ is a particular generating set of an ideal $I$ defined with respect to a monomial ordering.  One definition is that the ideal generated by the leading terms of $I$ is generated by the leading terms of $G$.  Gr\"obner Bases (GB) allow properties of the ideal to be deduced such as dimension and number of zeros and so are one of the main practical tools for working with polynomial systems.  Their properties and an algorithm to derive a GB for any ideal were introduced by Buchberger in his PhD thesis of 1965 \cite{Buchberger2006}.  There has been much research to improve and optimise GB calculation, with the $F_5$ algorithm \cite{Faugere2002} perhaps the most used approach currently. 

Like CAD the calculation of GB is necessarily doubly exponential in the worst case \cite{MM82} (when using a lexicographic order), although recent work in \cite{MR13} showed that rather than being doubly exponential with respect to the number of variables present the dependency is in fact on the dimension of the ideal.  Despite this worst case bound GB computation can often be done very quickly usually to the point of instantaneous for any problem tractable by CAD.

A reasonably common CAD technique is to precondition systems with multiple ECs by replacing the ECs by their GB.  I.e. let $E = \{e_1, e_2, \dots\}$ be a set of polynomials; 
$G = \{g_1, g_2, \dots\}$ a GB for $E$; and $B$ any Boolean combination of constraints, $f_i \, \sigma_i \, 0$, where $\sigma_i \in \{ <, >, \leq, \geq, \neq, =\}$) and $F = \{f_1, f_2, \dots\}$ is another set of polynomials.  Then 
\begin{align*}
\Phi &= (e_1 = 0 \land e_2 = 0 \land \dots) \land B \\
\Psi &= (g_1 = 0 \land g_2 = 0 \land \dots) \land B
\end{align*}
are equivalent and a CAD truth-invariant for either could be used to solve problems involving  $\Phi$.  

As discussed, the cost of computing the GB itself is minimal so the question is whether it is beneficial to CAD.  The first attempt to answer this question was given by Buchberger and Hong in 1991 \cite{BH91} (using GB and CAD implementations in the \textsc{SAC-2} system \cite{Collins1985}).  These experiments were carried out before the development of reduced projection operators and so the CADs computed were sign-invariant (and thus also truth-invariant for the formulae involved). Of the 10 problems studied: 6 were improved by the GB preconditioning, (speed-up from 2-fold to 1700-fold); 1 problem resulted in a 10-fold slow-down; 1 timed out after GB but completed without; and the other 2 were intractable both for CAD and GB.
The problem was recently revisited by Wilson et al. \cite{WBD12_GB} who studied the same problem set using \textsc{Qepcad-B} for the CAD and \textsc{Maple 16} for the GB.  There had been a huge improvement to the GB computation but it was still the case that two of the problems were hindered by GB preconditioning.  A recent machine learning experiment to decide when GB precondition should be applied \cite{HEDP16} found that 75\% of a data set of 1200 randomly generated CAD problems benefited from GB preconditioning.

If we consider GB preconditioning of CAD in the knowledge of the improved projection schemes for ECs (Subsection \ref{SUBSEC-ECProj}) then we see an additional benefit from the GB.  It provides ECs which are not in the main variable of the system removing the need for iterated resultants to find implicit ECs to use in subsequent projections. 

Since our aim is to produce one EC in each of the last $\ell$ variables, we need to choose an ordering on monomials which is lexicographic with respect to $x_n\succ x_{n-1}\succ\cdots\succ x_{n-\ell+1}$: it does not actually matter (from the point of view of the theory: general theory suggests that `total degree reverse lexicographic in the rest' would be most efficient in practice) how we tie-break after that.

Let us suppose (in line with \cite{EBD15}) that we have $\ell$ ECs $f_1,\ldots,f_{\ell}$ (at least one of then, say $f_1$ must include $x_n$, and similarly we can assume $f_2$ includes $x_{n-1}$ and so on), such that these imply (even over $\mathbb C$) that the last $\ell$ variables are determined (not necessarily uniquely) by the values of $x_1,\ldots,x_{n-\ell}$.  Then the polynomials $f_1$, $\Res_{x_n}(f_1,f_2)$, $\Res_{x_n,x_{n-1}}(f_1,f_2,f_3)$ etc. are all implied by the $f_i$. Hence either they are in the GB, or they are reduced to 0 by the GB, which implies that smaller polynomials are in the GB. Hence our GB will contain polynomials (which are ECs) of degree (in each variable separately) at most
\[
d, \, 4d^2, \, 27d^3, \, \dots, \, ((\ell+1)d)^{\ell+1}.
\]
Note that we are not making, and in the light of \cite{MR13} cannot make, any similar claim about the polynomials in fewer variables. Note also that it is vital that the equations be in the last variables for this use of \cite{Jouanolou1991,BM09} to work.

\section{Worked Example}
\label{SEC-Example}

We will work with the polynomials
\begin{align*}
f_1 &:= xy - z^2 - w^2, \qquad
f_2 := x+y^2+z+w, \\
f_3 &:= x-y^2+z-w, \qquad
h := x+y+z+w,
\end{align*}
and the semi-algebraic system
\[
\phi := f_1=0 \land f_2=0 \land f_3 = 0 \land h>0.
\]
We assume a variable ordering $z \succ y \succ x \succ w$ (meaning we will first project with respect to $z$) and seek a CAD truth-invariant for $\phi$.  

We have 3 ECs to take advantage of.  However, they all have main variable $z$ and so only one of them may be a designated EC for projection purposes.  The existing theory \cite{McCallum2001}, \cite{EBD15} would suggest propagating the ECs by calculating:
\begin{align*}
r_1 &:= \res(f_1, f_2, z) = -y^4 - 2wy^2 - 2xy^2 - 2w^2 - 2wx - x^2 + xy, \\
r_2 &:= \res(f_1, f_3, z) = -y^4 - 2wy^2 + 2xy^2 - 2w^2 + 2wx - x^2 + xy, \\
r_3 &:= \res(f_2, f_3, z) = -2y^2 - 2w;
\end{align*}
and then
\begin{align*}
R_1 &:= \res(r_1, r_2, y) = 256x^4(w^4+2w^2x^2+x^4+wx^2), \\
R_2 &:= \res(r_1, r_3, y) = \res(r_2, r_3, y) \\ \qquad &= 16w^4+32w^2x^2+16x^4+16wx^2.
\end{align*}
Of course, since $R_1$ contains $R_2$ as a factor we have $\res(R_1, R_2, x) = 0$.  

We have multiple choices for running Algorithm \ref{alg:ECProj} since we can only declare one polynomial as an EC with a set main variable.  There are hence $3 \times 3 \times 2 = 18$ possible configurations.  We have built the CAD for each choice (lifting using the improved procedure developed in \cite{EBD15}).  Depending on the choice made the number of cells in the final CAD varies between 73 and 335 (with average value 185).  The minimum is achieved using designations $(f_2, r_3, R_2)$.  

Now consider taking a GB of $\{ f_1, f_2, f_3 \}$.  We use a plex monomial ordering on the same variable ordering as the CAD to achieve a basis defined by 
\begin{align*}
g_1 &= z+x, \qquad
g_2 = y^2+w, \qquad
g_3 = -w^2-x^2+xy \\
g_4 &= w^2x+w^2y+x^3+wx, \qquad 
g_5 = x^4+2x^2w^2+w^4+x^2w
\end{align*}
This is an alternative generating set for the ideal defined by the explicit ECs and thus all the $g_i=0$ are also ECs for $\phi$.  We consider using these as the designated ECs when building the CAD.  There is no longer any choice regarding the EC with mvar $z$ or $x$ but there are 3 possibilities for the designation with mvar $y$.  Designating $g_2$ yields 73 cells (the best previous cell count) while either of $g_3$ or $g_4$ yields 45 cells.  Note in particular that the degrees of the GB polynomials are on average lower (and never greater) than those of the iterated resultants.

\section{Sketch of the effect on complexity}
\label{SEC-Complexity}

Following Section \ref{SEC-Deg} we see that when building a lexicographical basis the degree of the polynomials in the GB is restricted and thus this will be a better method for the identification of implicit ECs to use in subsequent projections than iterated resultants.  Let us sketch how this will effect the complexity of CAD following the techniques set out in \cite{BDEMW16}, \cite{EBD15} and summarised in Section \ref{SUBSEC-ECProj}.

The designated ECs will have lower degrees $d, 4d^2, 27d^3$ and in general $(sd)^{s}$ for the EC with mvar $x_{n-s}$.  From now on we will ignore the constant factors and focus on the exponents of $d$ generated.  This is both for simplicity in the analysis, and because we have not found a closed form solution for the product of the constant factors in the new analysis.  But we note that when using GB the constant factors grow exponentially in $\ell$ while with iterated resultants they grow doubly exponentially in $\ell$ (as in Table \ref{tab:P}).  Further, the constant term can be shown to be strictly lower for all but the first few projections, with the issue there a laxness of the analysis not the algorithm (as in Section \ref{SUBSEC-IR} we saw that the multivariate resultants was itself a factor of the iterated resultant).

We keep track of both the degree of the designated EC and the degree of the entire set of polynomials.  The reduced projection operator $P_F(B)$ will still take discriminants and coefficients of these; and resultants of them with the other projection polynomials.  Thus the highest degree polynomial produced grows with the exponent of $d$ being the sum of the exponent from the designated EC and that from the other polynomials.  
This generates the top half of Table \ref{tab:P2} and we see that the exponents form the so called \emph{Lazy Caterer's sequence}\footnote{The On-Line Encyclopedia of Integer Sequences, 2010, Sequence A000124, https://oeis.org/A000124} otherwise known as the Central Polygonal Numbers.  The remaining projections use the sign-invariant projection operator and so the degree is squared each time, leading to the bottom half of Table \ref{tab:P2}.

\begin{table}[ht]
\caption{Maximum degree of projection polynomials produced for CAD when using projection operator (\ref{eq:ECProjStar}) $\ell$ times and then (\ref{eq:P}).}  \label{tab:P2}
\begin{center}
\begin{tabular}{c@{\hskip 0.1in}|@{\hskip 0.1in}cc}
\multirow{2}{*}{\textbf{Variables}} & \multicolumn{2}{c}{\textbf{Maximum Degree}} \\
                           & \textbf{EC} & \textbf{Others}                \\
\hline
$n$          & $d$       & $d$            \\
$n-1$        & $d^2$     & $d^2$       \\
$n-2$        & $d^3$     & $d^4$     \\
$n-3$        & $d^4$     & $d^7$                 \\
\vdots       & \vdots    & \vdots   \\
$n-\ell$     & $d^{\ell+1}$ & $d^{\ell(\ell+1)(\nicefrac{1}{2}) + 1}$  \\
 & & \\
\hline
 & & \\
$n-(\ell+1)$ & \multicolumn{2}{c}{$d^{\ell(\ell+1) + 2}$}    \\
$n-(\ell+2)$ & \multicolumn{2}{c}{$d^{2\ell(\ell+1) + 2^2}$}    \\
$n-(\ell+3)$ & \multicolumn{2}{c}{$d^{2^2\ell(\ell+1) + 2^3}$}    \\
\vdots       & \multicolumn{2}{c}{\vdots}                  \\
$n-(\ell+r)$ & \multicolumn{2}{c}{$d^{2^{r-1}\ell(\ell+1) + 2^r}$} \\
\vdots       & \multicolumn{2}{c}{\vdots}         \\
1            & \multicolumn{2}{c}{$d^{2^{n-\ell-2}\ell(\ell+1) + 2^{n-\ell-1}}$}
\end{tabular}
\end{center}
\end{table}

We can now consider the generic bound (\ref{eq:BoundGeneral2}) using the degrees from Table \ref{tab:P2} as the $d_i$.  The term with base $d$ may be computed by
\[
\prod_{s=0}^{\ell} \big( d^{s+1} \big) \prod_{r=1}^{n-\ell-1} \big( d^{2^{r-1}\ell(\ell+1) + 2^{r}} \big).
\]
The exponent of $d$ evaluates to
\begin{equation}
\label{eq:newDegreeBound}
2^{(n-\ell)}\tfrac{1}{2}(\ell^2 + \ell + 2) - \tfrac{1}{2}(\ell^2 + \ell) - 2.
\end{equation}

Let us compare this with the term with base $m$ from (\ref{eq:BoundEC1}).  As with the improvements in \cite{EBD15}, the improvements here have allowed the reduction of the double exponent from by $\ell$, the number of ECs used.  However, the reduction is not quite as clean as the exponential term in the single exponent is multiplied by a quadratic in $\ell$.  This is to be expected as the singly exponential dependency on $\ell$ in the Number column of Table \ref{tab:P} was only in the term with constant base while for Table 2 the term with base $d$ is itself single exponential in $\ell$.

\newpage

\section{Discussion}
\label{SEC-Summary}

We have considered the issue of CAD in the presence of multiple ECs.  We followed our recent work in \cite{EBD15} which reduced the complexity with respect to the number of polynomials to see if similar achievements can be obtained with the respect to polynomial degree.  We suggest that using Gr\"{o}bner Bases in place of iterated resultants allows for the same reduction in the double exponent by the number of ECs used during projection. 

We have sketched the complexity results but defer the full analysis (using the (M,D)-property as detailed in \cite{BDEMW16}) until a number of issues can be cleared up.  We discuss some of them here:
\begin{itemize}
\item Will using GB not risk increasing the base number of polynomials in $m$? 
\end{itemize}
On one level this seems unlikely (since we are starting with a generating set all in the main variable and deriving another which would mostly not be) but we have yet to rule it out.  Of course, the number of polynomials in the input can bear little relation to the number generated by projection.

We note that there is an alternative way to use GB for CAD than that outlined in Section \ref{SUBSEC-GB} (replacing a set of ECs by another).  We could instead use the GB purely as an implicit EC generation tool; and just add selected polynomials from it to our input without replacing anything.  For example, the GB in the worked example of Section \ref{SEC-Example} had 3 polynomials with main variable $y$ only one of which can be the designated EC.  Rather than replacing all the $f_i$ by all the $g_i$ we could instead just add 2 of the $g_i$ (one in main variable $y$ and one in $x$) to the input set to act as designated ECs at lower levels.  This approach would cap the increase in $m$ to the number of designated ECs we can identify.
Taking this approach with the example in Section \ref{SEC-Example} would not change the final cell counts.

\begin{itemize}
\item Will the GB always produce as many ECs with different main variables as the iterated resultant method?
\item How to proceed in the case where we have non-primitive ECs?  
\end{itemize}
This paper (as with most previous work on ECs) only deals with the case of primitive designated ECs.  We refer the reader to the final section of \cite{EBD15} where we sketch approaches that could be adapted to deal with this (including the theory of TTICAD from \cite{BDEMW13} \cite{BDEMW16}).
\begin{itemize}
\item How is the complexity affected when the projections using ECs are not in strict succession?
\item Can we mix the orderings in the CAD and the GB?
\end{itemize}

Finally, we return to the fact acknowledged in Section \ref{SUBSEC-GB} that previous work on using GB to precondition CAD \cite{BH91}, \cite{WBD12_GB}, \cite{HEDP16} has found that it is not always beneficial and how that interacts with the claims of this paper.  The simple answer is that the analysis offered here is of the worst case and makes no claim to the average complexity.  However, we actually hypothesise that it was it was the fact that the CAD computations involved in those paper did not take advantage of the new multiple EC technology which will account for many of the cases were GB hindered CAD.  We plan future experiments to test this hypothesis.

\subsubsection*{Acknowledgements}

This work was originally supported by EPSRC grant: EP/J003247/1 and is now supported by EU H2020-FETOPEN-2016-2017-CSA project $\mathcal{SC}^2$ (712689).  
We are also grateful to Professor Buchberger for reminding JHD that Gr\"obner bases were applicable here.

\bibliographystyle{plain}
\bibliography{CAD}

\end{document}